\pdfoutput=1

\documentclass[%
 preprint,
 amsmath,amssymb,
 aps,
pra,
]{revtex4-2}

\usepackage{graphicx}% Include figure files
\usepackage{dcolumn}% Align table columns on decimal point
\usepackage{bm}% bold math
\usepackage{setspace}
\usepackage[bookmarks=false]{hyperref}
\begin{document}

\preprint{APS/123-QED}

\title{A Theoretical Paradigm for Thermal Rectification via Phonon Filtering and Energy Carrier Confinement}% Force line breaks with \\
\thanks{This work is currently under review.}%

\author{Brian F. Donovan}
 \email{bdonovan@usna.edu}
\affiliation{%
 Physics Department\\
 United States Naval Academy\\
 Annapolis, MD 21402}%

\author{Ronald J. Warzoha}
\affiliation{
 Department of Mechanical Engineering\\
 United States Naval Academy\\ 
 Annapolis, MD 21402% with \\
}%

\date{\today}% It is always \today, today,
\setstretch{1.5}
\begin{abstract}
{\bf We provide a theoretical framework for the development of a solid-state thermal rectifier through a confinement in the available population of phonons on one side of an asymmetrically graded film stack. Using a modification of the phonon gas model to account for phonon filtering and population confinement, we demonstrate that for an ideal material, with low phonon anharmonicity, significant thermal rectification can be achieved even in the absence of ballistic phonon transport. This formalism is used to illustrate thermal rectification in a thin-film of diamond (1-5 nm) graded to dimensions $>$ 1 $\mu$m exhibiting theoretical values of thermal rectification ratios between 0.75 and 6. Our theoretical formulation for thermal rectification is therefore expected to produce opportunities to design advanced solid-state devices that enable a variety of critical technologies.}

%\begin{description}
%\end{description}
\end{abstract}

\maketitle

%\tableofcontents

%w\section{\label{sec:level1} Introduction}
\noindent \textit{Introduction.--}The development of solid-state thermal architectures is expected to result in transformative technological breakthroughs similar to those realized in the information technologies sector. For instance, thin-film thermal rectifiers have the capacity to revolutionize phonons as information carriers, allowing for the materialization of phononic computing \cite{wang2007thermal}. Similarly, thermal biasing is pivotal for improvements in thermal barrier coating effectiveness and heat mitigation in electronic devices \cite{dames2009solid}. Problematically, such biasing has traditionally been achieved either when thermal gradients are sufficiently large to produce corresponding gradations in temperature-dependent thermal conductivity across a set of dissimilar materials \cite{dames2009solid, li2015temperature, hu2008thermal,roberts2011phonon,zhu2014temperature,wu2007thermal} {\it or} when there exists mass gradation in the direction of heat flow \cite{roberts2011review}.

Recent work by Chang et al. \cite{chang2006solid} describes a process to produce thermal rectification via the formation of an asymmetric structure that induces asymmetric boundary scattering of phonons with respect to the direction of heat flow. In their work, the authors find that a partial mass loading of asymmetrically deposited, amorphous C$_9$H$_{16}$Pt particles on the outer surface of a boron nitride nanotube results in a measured thermal rectification of $\sim$ 7$\%$. Others have also found asymmetry to be an effective mechanism to achieve thermal rectification \cite{yang2009thermal, pereira2011sufficient,yu2017investigation,zhong2011thermal,yang2007thermal,hu2008thermal,wang2014phonon}. Still, the magnitude of thermal rectification achieved through asymmetric structuring remains relatively low (i.e. $<$ 10-50$\%$) in the absence of significant thermal gradients in the direction of heat flow \cite{yang2009thermal}. In fact, the authors are aware of only one study that experimentally demonstrates a thermal rectification ratio well above this \cite{martinez2015rectification} by tunnel-coupling metals to superconducting elements. We note, however, the difficulty associated with the integration of such a device into practical thermal applications. In this work, we provide a physical construct that can be used to design thermal rectifiers that do not rely on thermal and/or mass gradients to produce an observable thermal rectification effect.%

To this end, we focus on the conditions necessary for thermal rectification from the perspective of available wavelengths in the phonon density of states as thermal energy traverses thin-film material stacks in opposing directions. 

\vspace{0.5cm}
%\section{\label{sec:level1} Modeling thermal rectification via phonon confinement}
\noindent \textit{Modeling thermal rectification via phonon confinement.--} To demonstrate thermal rectification in thin-film, asymmetric material stacks, we employ the analytical treatment of phonon transport constructed by Callaway \cite{callaway1959model} and represented by,

\begin{equation}
    \kappa = \frac{1}{3}\sum_{j}\int_{k}C_v\nu ldk
\end{equation}

and,

\begin{equation}
    C_v = \frac{1}{2\pi^2}\sum_{j}\int_{k}\hbar\omega\frac{\partial f_{BE}}{\partial T}k^{2}dk
\end{equation}

\noindent where $\kappa$ and C$_v$ are thermal conductivity and volumetric heat capacity, respectively. In the above expressions, j represents the polarization index, $\nu$ the phonon group velocity, $\omega$ the angular frequency, k the wavevector and $\frac{\partial f_{BE}}{\partial T}$ the temperature derivative of the Bose-Einstein distribution. As well, the phonon mean free path, $l$ is incorporated into the physical representation of thermal conductivity to account for energy carrier scattering. 

The impact of extrinsic scattering mechanisms on the spectrum of phonon wavelengths that contribute to the thermal conductivity is analyzed using Matthiessen's rule, which allows us to isolate the intrinsic phonon mean free path, $l_{in}$, and the mean free path considering the sample boundaries, $l_{bound}$. This results in a representative thermal conductivity for a given phonon polarization as described by, 

\begin{equation}
    \kappa = \frac{1}{3} \int C_{v} \nu \big[l_{in}^{-1} + l_{bound}^{-1}\big]^{-1}dk
    \label{kappa}
\end{equation}

\noindent In this work, we limit our analysis to one-dimensional heat flow across thin film configurations. Thus, l$_{bound}$ is treated as the characteristic length of an individual film.

Within the framework of spectral contributions to thermal conductivity, we deduce that thermal rectification occurs when long wavelength phonons are unable to carry thermal energy across the film in the presence of a very thin, specular material boundary on one side of the film. Thus we rewrite the equation for the thermal conductivity (Eqn. \ref{Kappa}) to highlight its wavelength ($\lambda$) dependence, and explicitly identify the components that are a function of phonon wave vector ($k=2\pi/\lambda$). Moreover, it will become informative to discuss impacts to the thermal conductivity within the context of the phonon density of states, DOS$(k)$, thus the integral that determines the thermal conductivity for a given phonon polarization as a function of wave vector can be written as, 

\begin{equation} \label{Kappa}
    \kappa(k)= \int_{k_{min}}^{k_{max}} \hbar \omega(k) \mathrm{DOS}(k) \frac{\partial f_{BE}}{\partial T} \nu(k) \big[l_{in}(k)^{-1} + l_{bound}^{-1}\big]^{-1} dk
\end{equation}

\noindent This integral is taken over the available wave vector space from $k_{min}=\pi/d$ to $k_{max}=\pi/a$, where $d$ is the physical dimension of the entire crystal and $a$ is the lattice constant of the crystal. In the limit of most bulk formulations of thermal conductivity, the minimum wave vector is assumed to be zero (i.e. an infinite crystal). 

\begin{figure}
    \centering
    \includegraphics[width=15cm]{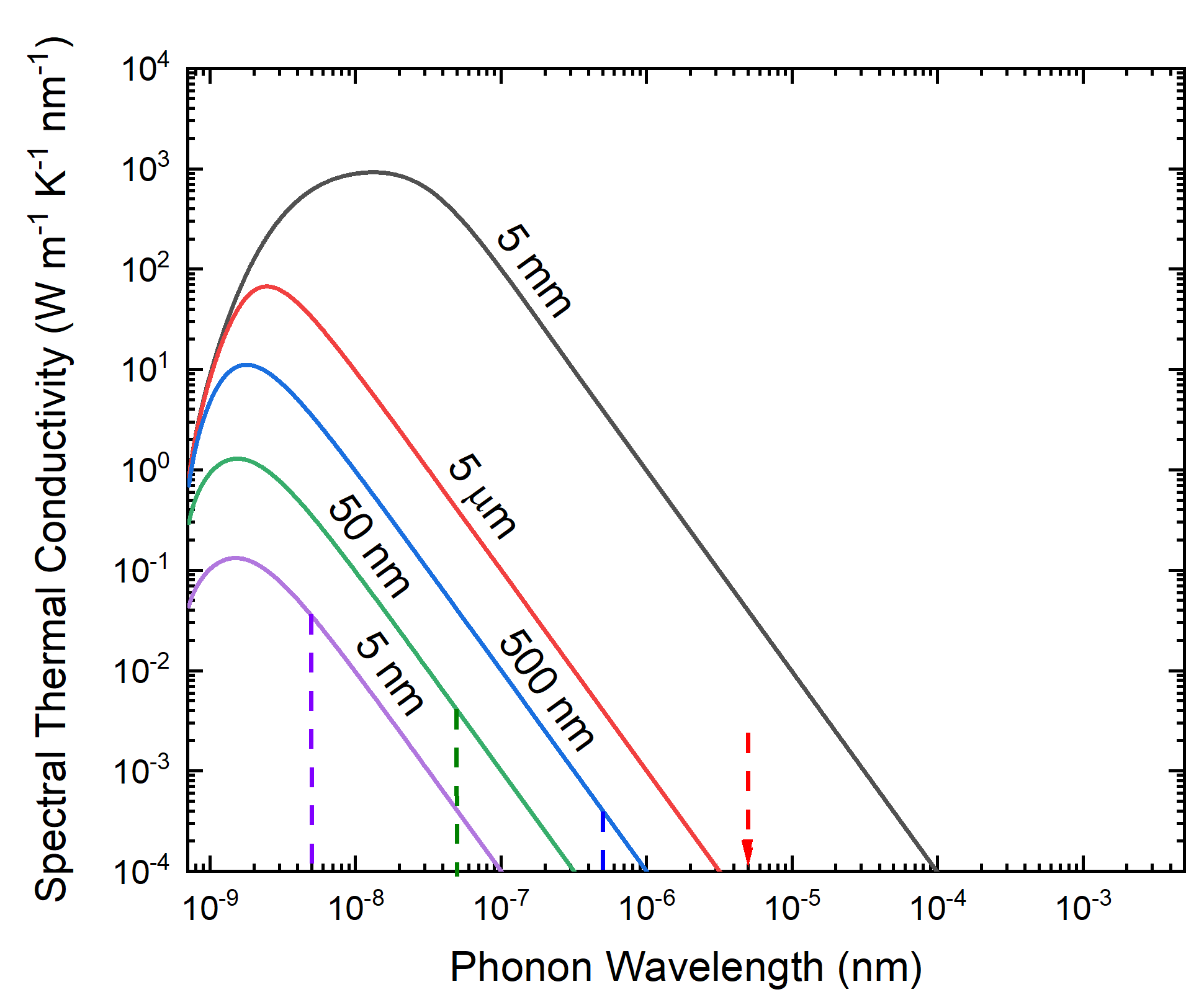}
    \caption{\bf \setstretch{1.0} Spectral thermal conductivity as a function of phonon wavelength from a typical scattering limited thin film model. This function is integrated to obtain the thermal conductivity for a given material system. Curves are included for diamond films with thicknesses of 5 mm (black), 5$\mu$m (red), 500 nm (blue), 50 nm (green) and 5 nm (purple). Also included are dashed lines indicating where the wavelength of phonons corresponding to the film thickness lies. Phonons with wavelengths greater than the film thickness contribute far less (orders of magnitude) than the primary heat carrying phonons in each system.}
    \label{SpecK}
\end{figure}

Typically, the impact of nanostructuring on thermal conductivity is assumed to be accounted for entirely by modifications to the phonon mean free path. In this case, phonons that have wavelengths greater than the dimensions of a given nanostructure (a material layer having nanometer thickness) are assumed to be present when calculating thermal conductivity using the standard model, but are thought to heavily scatter at any material boundary (i.e. grain boundaries, film boundaries, or other interfaces). It is useful in this context to consider spectral contributions to the thermal conductivity. In Fig. \ref{SpecK}, we observe the spectral thermal conductivity of the longitudinal acoustic phonon branch of diamond, which is calculated from the integrand of Eq. \ref{Kappa}, plotted against phonon wavelength for an infinite crystal ($d\rightarrow \infty$) with finite scattering dimensions of 5 mm. Further, we include the spectral thermal conductivity for 5 $\mu$m, 500 nm, 50 nm, and 5 nm films of diamond modeled using this same full integral for an infinite crystal, and account for the impacts of nanostructuring on thermal conductivity using a mean free path that is limited by boundary scattering. 

Figure \ref{SpecK} provides the relative contribution to the total thermal conductivity of single-layer diamond films with varying thickness at any given phonon wavelength. The Callaway model can be used to determine the effect that a given population of phonons has on total thermal conductivity by integrating over all possible phonon wavelengths. This has been addressed within the wider literature, most notably through the concept of mean free path accumulation \cite{donovan2014spectral,regner2015advances,cuffe2015reconstructing,jiang2016role,minnich2012determining,jain2013phonon}. One immediate conclusion that can be drawn from this analysis is that the peak phonon wavelength is significantly smaller than the film's characteristic dimension, and is further reduced as the sample dimension shrinks. 

We note now that this peak phonon wavelength contributes orders of magnitude more to the thermal conductivity than wavelengths that are on the order of the film thickness. To illustrate this point, the characteristic dimension used to determine the phonon mean free path by boundary scattering is plotted in Fig. \ref{SpecK} with a dashed, vertical line for each film thickness (or the bounds of the figure for the 5 mm film). When examining phonon wavelengths that extend beyond the limit of the physical boundary, their inclusion in the computation of film thermal conductivity can be omitted due to their relatively low contributions to thermal conductivity when compared to those wavelengths that span the peak. In the context of {\it phonon filtering}, boundary scattering preferentially eliminates (or filters out) contributions to the thermal conductivity from phonons that are greater than the film thickness. 

As a consequence of the relatively low contribution to the thermal conductivity made by phonons having wavelengths greater than the film thickness, we assert that these dimensions can be considered as the predominant confining features of the material. In fact, in a free standing nanostructured system (a suspended nanoparticle or unsupported membrane), the confinement of available phonons is clearly reduced to the nanostructured dimension and incorporation of longer wavelength phonons that are even available to undergo scattering would be {\it non-physical} \cite{ghosh2010dimensional,lindsay2010flexural}. 

In this work, we examine the impact of treating those boundaries, which have previously been considered only as scattering sites, as sites that confine the population of phonons available to transfer thermal energy across a layered material system. This implies that the lower bound of our integration in wave vector space is governed by the characteristic dimension of the nanostructured component. In the case of a spherical nanoparticle, for example, the relevant dimension is the diameter ($d$) of the nanoparticle. Likewise, for a free standing film with one dimensional heat flow, the film thickness ($t$) serves as the relevant dimension for phonon confinement. 

In order to account for possible phonon confinement effects, we modify the bounds of our integration such that we disregard wave vectors below $k_{min}=\pi/t$. Thus, every component of the thermal conductivity in Eq. \ref{Kappa} that depends on the wave vector will assume the effects of phonon filtering or dimensional phonon confinement. This simple modification provides for a more rigorous treatment of the physics that govern thermal transport in multi-layer nanoscale films, and allows for an additional mechanism to achieve thermal rectification in multi-layer material systems. 

\begin{figure}
    \centering
    \includegraphics[width=15cm]{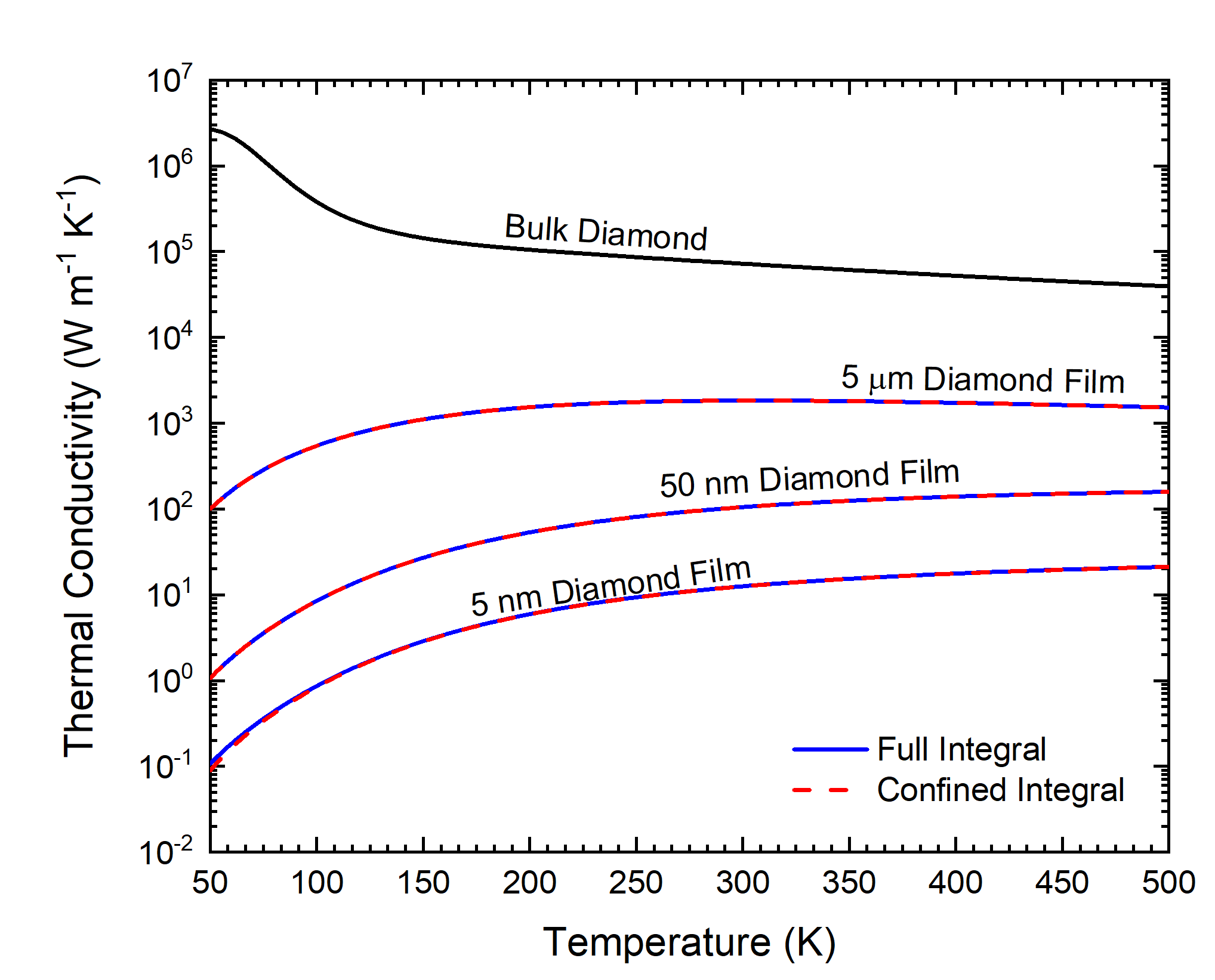}
    \caption{\bf \setstretch{1.0} Thermal conductivity of diamond modeled over temperature using both a standard scattering limited Callaway model as well as a confined phonon population model. Models are shown for samples with bulk dimensions (black, solid line) as well as films of 5 $\mu$m, 50 nm, and 5 nm. In the thin film models, the standard ``Full Integral'' scattering limited approach (blue line) matches the confined phonon population approach (red dashes) very well. The only deviation comes at low temperatures with the single-nanometer-scale film.}
    \label{KvsT}
\end{figure}

We validate our confined phonon model by first comparing computations of temperature-dependent thermal conductivity for single-layer diamond films to those obtained by integrating across all phonon wavelengths, as shown in Fig. \ref{KvsT}. In this case we use the acoustic branches of a real diamond phonon dispersion to determine the phonon frequency and group velocity \cite{warren1967lattice,slack1964thermal} and assume a spherical density of states. The intrinsic mean free path is determined by accounting for phonon-phonon scattering and phonon-impurity scattering when fit to literature data for bulk diamond systems. Values for bulk diamond thermal conductivity computed with the confined phonon integral are found to be consistent with those values available in the wider literature \cite{ma2014examining,donovan2014spectral,cheng2018thermal,cheng2018probing}.

In Fig.~\ref{fig:KvsT}, the confined phonon integration matches the full integral almost identically. The only deviation occurs at low temperature with single-digit nanometer sized films, which is physically appropriate considering the phonon population is not yet fully occupied in such thin films and long wavelength phonons begin to govern thermal transport. The comparison in Fig. \ref{KvsT} lends confidence that a confined phonon model captures phonon filtering identically to the Callaway model integrated over a full spectrum of all possible wavelengths.

\vspace{0.5cm}
\noindent \textit{Thermal rectification in nanostructured film stack.--}In order to use a phonon confinement effect to achieve thermal rectification, layered (or graded) structures must be fabricated in such a way that a confined phonon population can be injected from one layer into an otherwise non-confined layer. Within a practical context, an ultra-thin film can be deposited above a thick film (whose phonon population is unconfined) to achieve this effect. Here, the ultra-thin film acts to filter out long wavelength phonons, where the available phonon population across the film boundary remains limited to the spectrum of incoming phonons. 

For the available phonon population to remain consistent on opposing sides of the film boundary, the interface between the phonon filter layer and adjacent thick layer(s) must be lattice matched and coherent. If the morphology of the interface results in diffuse and anharmonic phonon scattering, then the transport of thermal energy from one film to the other would result in a redistribution of phonons that assumes the thick film's enlarged density of states, rendering this formulation invalid. 

Likewise, the manifestation of thermal rectification requires a lack of redistribution of the phonon population from anharmonic phonon-phonon scattering within the thick layer itself. While we account for these effects to determine the intrinsic mean free path within our model, we assume that the phonons injected into the thick film from the filter layer do not scatter anharmonically into other wavelengths that may not have been available to begin with. Each of these requirements are fulfilled in several existing material systems, including grain-graded, nanocrystalline diamond membranes (whose grain boundaries are twinned) \cite{sood2016anisotropic} and Si-based superlattices \cite{fan2001sigec}.

\begin{figure}
    \centering
    \includegraphics[width=17cm]{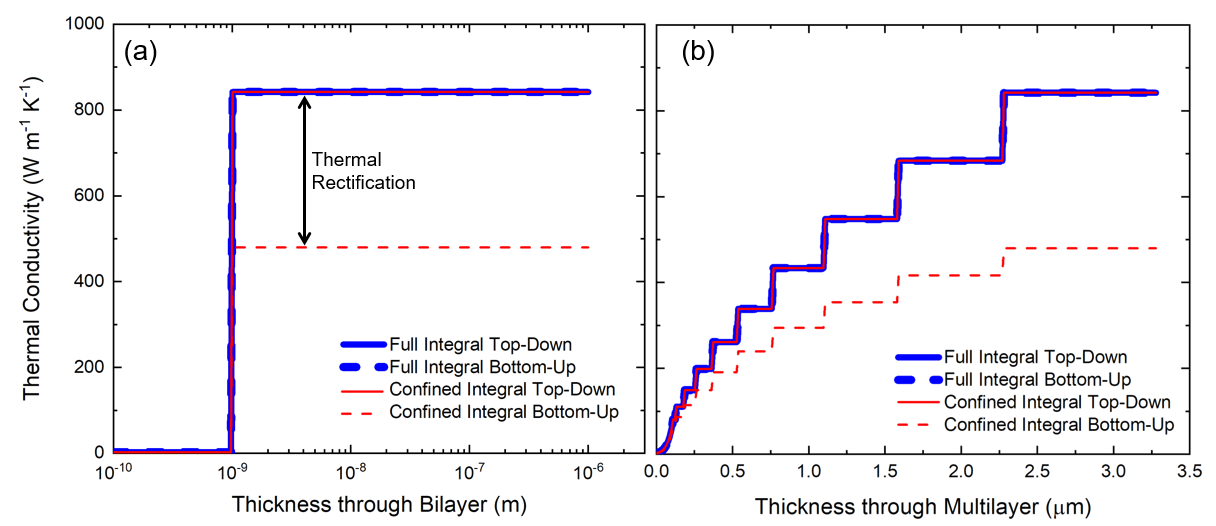}
    \caption{\bf \setstretch{1.0} Thermal conductivity of diamond through the thickness of a bilayer (a) and multilayer (b) film going from the bottom side of the total film stack up (thin to thick) and the topside down (thick to thin). Models are shown using only scattering limited modeling (blue solid line and dashes, no directional difference) and the confined phonon population model (red dashed line from bottom up and solid line from top down). In the top-down direction both modeling approaches match, however in the bottom up, the confined phonon population model results in a limited thick-film thermal conductivity due to a lack of long wavelength phonons present in the incoming phonon population. This results in a significant difference in thermal conductivity and a large thermal rectification effect. }
    \label{KvsLayer}
\end{figure}

Provided the above requirements are met, this treatment can be used to design a solid-state thermal rectifier using multilayer thin-films. The most palpable system to imagine is a bilayer stack having one ultra-thin layer deposited above a thick layer. In Fig. \ref{KvsLayer}a we provide an example of thermal rectification in a bilayer diamond film system that consists of a 1 $\mu$m film above a 1 nm film. Fig. \ref{KvsLayer} provides for an interesting result. In the typical ``full integral'' model, we see that the thermal conductivity is not directionally dependent when calculated using mean free paths that are limited by boundary scattering (as expected). Conversely, the phonon confinement analysis {\it does} yield a directionally-dependent thermal conductivity. When we apply these physics to the case when heat flows from the thick film to the thin film, the confined thermal model results in a thermal conductivity distribution identical to that obtained from the ``full integral'' model. However, if heat flows in the other direction, we obtain a significantly reduced thermal conductivity in the thick layer as we are not fully populating all available phonon modes. The directional-dependence of thermal conductivity results in thermal rectification, and is extremely significant in the limit of full phonon confinement.

Mathematically, the resulting thermal rectification originates from modifications to the limits of integration in Eq. \ref{Kappa}. Considering the case when heat emanates from the thick-film side, $k_{min}={\pi}~{\mu\mathrm{m}^{-1}}$ (as well as $l_{bound}=1~\mu\mathrm{m}$), which then become limited in the thin film portion of the stack. In the opposing direction, however, $k_{min}=\pi~\mathrm{nm}^{-1}$ (and $l_{bound}=1~\mathrm{nm}$). Even though the mean free path is relaxed as the heat moves into the thick layer ($l_{bound}=1~\mu\mathrm{m}$), $k_{min}$ does not change and so the confinement results in diminished thermal transport. 

In Fig. \ref{KvsLayer}b we also examine these physics for a 20 layer material stack with logarithmically increasing thickness (in this case, ranging from 1 nm to 1 $\mu$m). This is the type of dimensional evolution that one might see from nucleation and grain growth in a typical top-down film deposition technique \cite{sood2016anisotropic}. Here, our confinement still originates from the initial filtering layer (1 nm). We again note that this model assumes that no anharmonic interactions occur, however, traversing this many interfaces in a real material system would inevitably lead to {some} redistribution of phonon populations. Nevertheless, this provides an upper limit to thermal rectification (in the absence of any external temperature gradient or material asymmetry).

A thermal rectification ratio, $TR$, is used to calculate the degree of thermal rectification achieved relative to other works \cite{yu2017investigation,roberts2011review} and is represented by,

\begin{equation}\label{TR}
    TR=\frac{\kappa_{TD}-\kappa_{BU}}{\kappa_{BU}}
\end{equation}

\noindent where $\kappa_{TD}$ is the thermal conductivity with heat moving from the top down and $\kappa_{BU}$ is the thermal conductivity from the bottom up. 

\begin{figure}
    \centering
    \includegraphics[width=15cm]{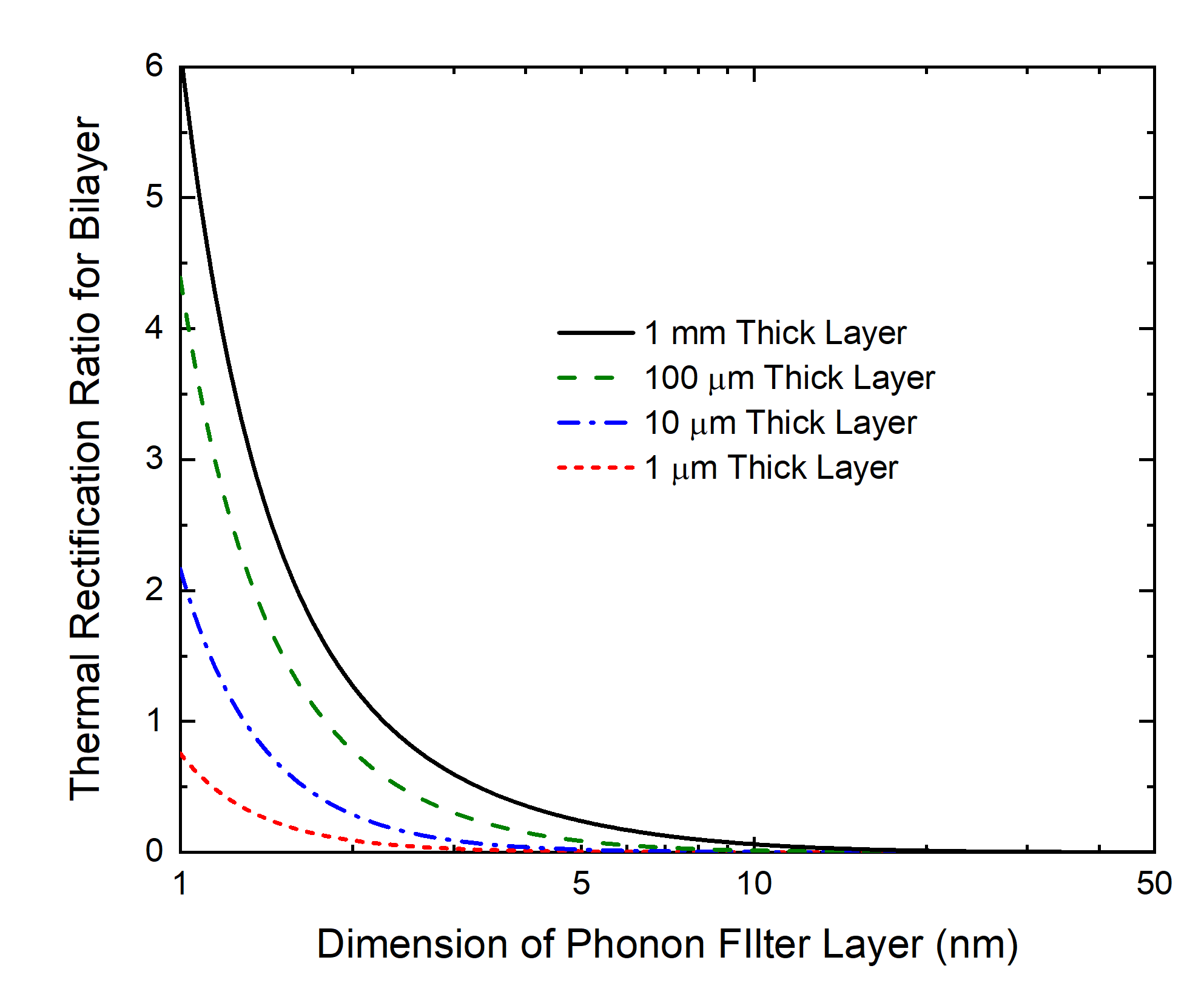}
    \caption{\bf \setstretch{1.0} Thermal rectification over various phonon filter dimensions (i.e. the thin layer in the bilayer film stack) computed for diamond films using the confined phonon population model. This is shown for multiple dimensions of the thick film, where thermal rectification effects become more extreme with a greater disparity between the filter layer and the top layer. Rectification is computed using the directional difference in the thermal conductivities in the top layer.}
    \label{TR1}
\end{figure}

The thermal rectification ratio is plotted for the bilayer system over various phonon filtering layer thicknesses in Fig. \ref{TR1}. In this formulation we compare the directional thermal conductivities in the thick layer as it represents the largest thermal resistor in the bilayer stack. We also include distributions for a number of different thick layer length scales, which reveal that despite primary limitations of the phonon filter layer, boundary scattering will still play a role in the thick layer's available phonon population. Effectively, thick layers with less boundary scattering are at greater risk for diminished thermal conductivity due to phonon confinement. However, even in a bilayer that consists of a 1 $\mu$m film with a 1 nm filter layer, the thermal rectification ratio is $>$ 75$\%$. We note that this is considerably higher than thermal rectification ratios that have been reported in literature (for cases where a large temperature gradient across the structure is absent). In the case where there is significant dimensional mismatch, extremely large thermal rectification ratios can be achieved ($>$ 600\%). The upper range of these thermal rectification ratios is expected to revolutionize a wide range of thermal devices and facilitate the development of phononic computing.

Figure \ref{TR1} suggests that when the filter layer thickness is greater than $\sim$10 nm, thermal rectification is negligible. This can be readily understood via the spectral contributions to thermal conductivity shown in Fig. \ref{SpecK}. Here, the peak contributors to the thermal conductivity exist in the sub-10 nm range (particularly in the thin film regime). We should note that while the bilayer system has a rectification factor that is governed by the thicker component, the multilayer system does not have the same spatial response. When multiple layers are present, the thermal rectification ratio is computed using average directional thermal conductivities in Eq. \ref{TR}. This results in a thermal rectification ratio that is slightly reduced (63\% in the 20 layer system that goes from 1 nm to 1 $\mu$m compared to 75\% in the same extremes for a  bilayer). 

\vspace{0.5cm}
\noindent \textit{Discussion.--} The use of a phonon filter layer to confine a population of phonons across a thin-film boundary is extremely promising for the realization of effective thermal rectifiers. These physical dynamics enable thermal rectification at realistic device scales with sample configurations that readily lend themselves to microelectronics processing techniques. Critically, extreme thermal rectification ratios of several hundred percent are predicted for these systems, which will finally allow for robust experimental demonstrations of thermal rectification.

We have enabled this physical understanding by modifying the well-known Callaway formulation for the thermal conductivity of a material based on heat capacity and thermal carrier dynamics. Critical to realizing thermal rectification, we have extended the model by considering the confinement of phonons when film thickness is sufficiently small. By restricting the integration limits to the characteristic dimension(s) of a film, we can directly observe the impact of phonon confinement on the density of phonon states available for thermal transport while simultaneously capturing the phonon filtering effects that are well-known within the scientific community. This can be extended to multilayer, graded materials, which are an important class of materials for integration into larger system platforms.

Though this analysis relies on a lack of anharmonic interactions across interfaces and within each material layer, it does provide an upper limit to thermal rectification that can be achieved in an ideal system. In diamond (a material that does, in fact, demonstrate negligible anharmonic scattering \cite{warren1967lattice}), this upper limit is computed to be $\sim$600\%, which is orders of magnitude larger than any experiments yet to be reported in literature. This is particularly significant given the additional potential to combine this effect with other mechanisms that are known to produce thermal rectification, such as geometric asymmetry and an imposed thermal gradient. Even if a fraction of this limit should be realized experimentally, it would allow for the production of practical thermal devices and represent a major advancement in the thermal sciences. 

\vspace{0.5cm}
\noindent \textit{Acknowledgements.--} BFD and RJW would like to acknowledge the financial support of Mr. Peter Morrison and the Office of Naval Research under Contract No. N0001419WX00501. The authors are also grateful for critical insight provided by Dr. Andrew Smith.

\bibliography{apssamp}% Produces the bibliography via BibTeX.

\end{document}